\documentclass{elsart}
\usepackage{epsfig}

\def\url#1{{\ttfamily\def\/{/\discretionary{}{}{}}#1}}

\def\ltsima{$\; \buildrel < \over \sim \;$}
\def\simlt{\lower.5ex\hbox{\ltsima}}
\def\gtsima{$\; \buildrel > \over \sim \;$}
\def\simgt{\lower.5ex\hbox{\gtsima}}

\begin{document}

\begin{frontmatter} 
\title{X--ray spectra transmitted through Compton--thick absorbers} 
\author{Giorgio Matt, Fulvio Pompilio \& Fabio La Franca\thanksref{email}}
\address{Dipartimento di Fisica, Universit\`a degli Studi ``Roma Tre",
Via della Vasca Navale 84, I--00146 Roma, Italy } 
 \thanks[email]{E-mails: (matt,pompilio,lafranca)@fis.uniroma3.it }
\begin{abstract} 
X--ray spectra transmitted through matter which is optically thick
to Compton scattering are computed by means of Monte Carlo simulations.
Applications to the BeppoSAX data of the Seyfert 2 galaxy in Circinus, 
and to the
spectral modeling of the Cosmic X--ray Background, are discussed.
\end{abstract} 
 

\end{frontmatter} 

\section{Introduction}

The presence of large amount of ``cold" (i.e. not much ionized, and
substantially opaque in X--rays)
matter around Active Galactic Nuclei is now a well
established fact. For all Seyfert 2 galaxies observed in X--rays so far there 
is evidence for absorption in excess of the Galactic one. In a 
significant fraction of them, the column density
of the absorbing matter exceeds 10$^{24}$ cm$^{-2}$ (Maiolino et al. 1998),
and the matter is therefore optically thick to Compton scattering. In a few
objects, like NGC~1068 (Matt et al. 1997), 
the column density is so high that the X--ray photons cannot 
escape even in hard X--rays, 
being trapped in the matter, downscattered to energies where photoelectric
absorption dominates, and eventually destroyed.
In other cases, like NGC~4945 (Iwasawa et al. 1993;
Done et al. 1996), Mrk~3 (Cappi et al. 1999) and the Circinus Galaxy
(Matt et al. 1999), the column density is a few$\times$10$^{24}$ 
cm$^{-2}$, so permitting the transmission of
a significant fraction of X--ray photons, many of them escaping
after one or more scatterings.
To properly model transmission through absorbers with 
these intermediate column densities, is therefore necessary to fully
take into account Compton scattering. This has been done in an analytical,
approximated way by Yaqoob (1997), but his model is valid only below 
$\sim$15 keV, a painful limitation after the launch of BeppoSAX, which 
carries a sensitive hard X--ray (15-200 keV) instrument, and in view
of future missions like Astro-E and Constellation-X. 

We have therefore calculated transmitted spectra
by means of Monte Carlo simulations. The code is, as far as
physical processes are concerned, basically that 
described in Matt, Perola \& Piro (1991). 
A spherical geometry, with the X--ray source in the centre, has been 
assumed, while the element 
abundances are those tabulated in Morrison \& McCammon (1983).
Photoelectric absorption, Compton scattering (in a fully relativistic
treatment) and fluorescence (for iron atoms only) are included in the code.
Photon's path are followed until either the photon is photoabsorbed
(and not re--emitted as iron fluorescence) or escapes from the cloud. 

Spectra have been calculated for 31 different column densities, ranging
from 10$^{22}$ to 4$\times10^{24}$ cm$^{-2}$.
In order to be independent of the shape of the primary radiation,
transmitted spectra for monochromatic emission have been calculated, 
with a step
of 0.1 keV below 20 keV, and 1 keV above. A grid has then been constructed,
which can be folded with the chosen spectral shape. 


\section{Transmitted spectra. Comparison with simple absorption models}

To illustrate the effects of including the Compton scattering
in the transmission spectrum, we show in Figs 1-5 (which refer to column
densities of 10$^{23}$, 3$\times10^{23}$, 
10$^{24}$, 3$\times10^{24}$ and 10$^{25}$ cm$^{-2}$, respectively), 
the results of the MonteCarlo
simulations (solid lines), along with the results when only photoelectric
absorption (dotted lines) or both photoelectric and Compton absorption
without scattering (dashed lines) are included. The first case is unphysical,
and it is shown here only for the sake of illustration; the second case would
correspond to absorption by matter with a negligible
covering factor to the primary source (i.e. a small cloud on 
the line of sight), a physically possible but highly unlikely situation,
at least for Seyfert galaxies (the fraction of Compton--thick sources
is estimated to be at least 30\%, Maiolino et al. 1998, and then the
covering fraction of the matter must be significant).
The injected spectrum is a power law with a photon index of 2 
and an exponential cut--off at 500 keV, as typical for Seyfert galaxies
(even if the latter parameter is at present poorly known).
As expected, our curves lie below the dotted curves
(because of the larger absorption especially above $\sim$10 keV, where
the Compton cross section starts dominating over the photoelectric cross
section), and above the dashed curves (because of the extra radiation
provided by the scattering of photons into the line of sight).
The effect is dramatic, especially for large column densities and
high energies. 

\begin{figure}
\epsfig{file=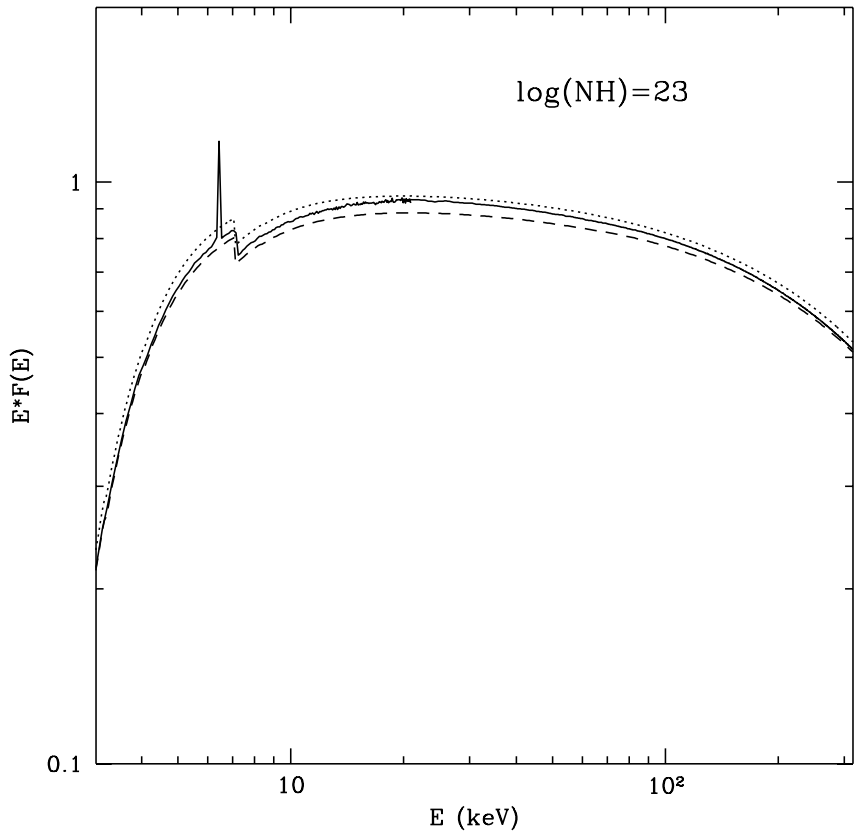, height=12.cm, width=12.cm, angle=0.}
\caption{Transmitted spectra (in $EF(E)$) for a column density 
N$_{\rm H}$=$10^{23}$
cm$^{-2}$. The solid line refer to the Monte Carlo results discussed in this
paper. The dotted line is for photoelectric absorption alone (an unphysical 
situation, presented here 
only for the sake of comparison); the dashed line for
complete (photoelectric and Compton scattering) absorption, as for a 
small cloud on the line of sight.}
\end{figure}

\begin{figure}
\epsfig{file=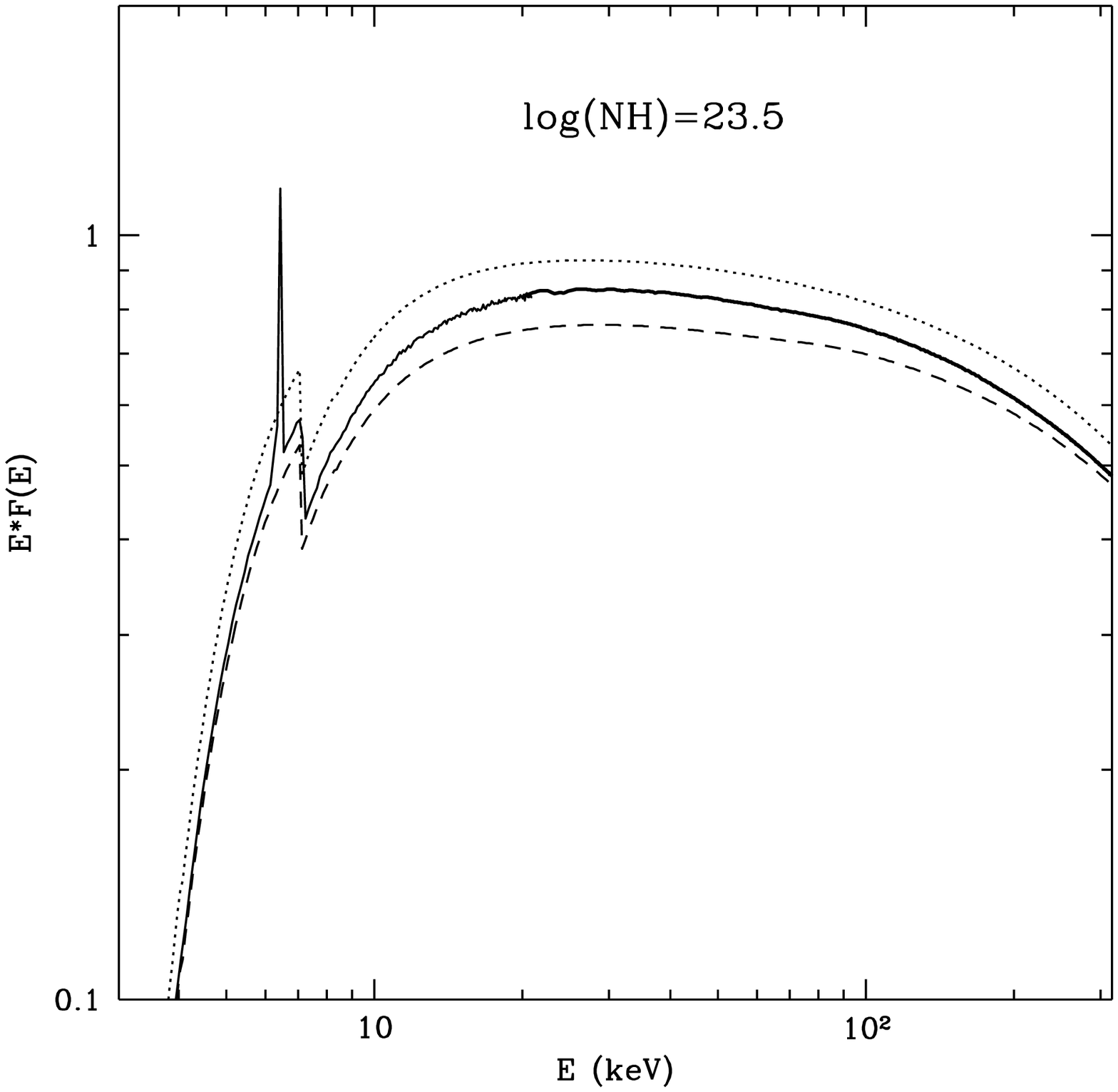, height=12.cm}
\caption{ As for Figure 1, but for N$_{\rm H}$=$3\times10^{23}$
cm$^{-2}$.}
\end{figure}

\begin{figure}
\epsfig{file=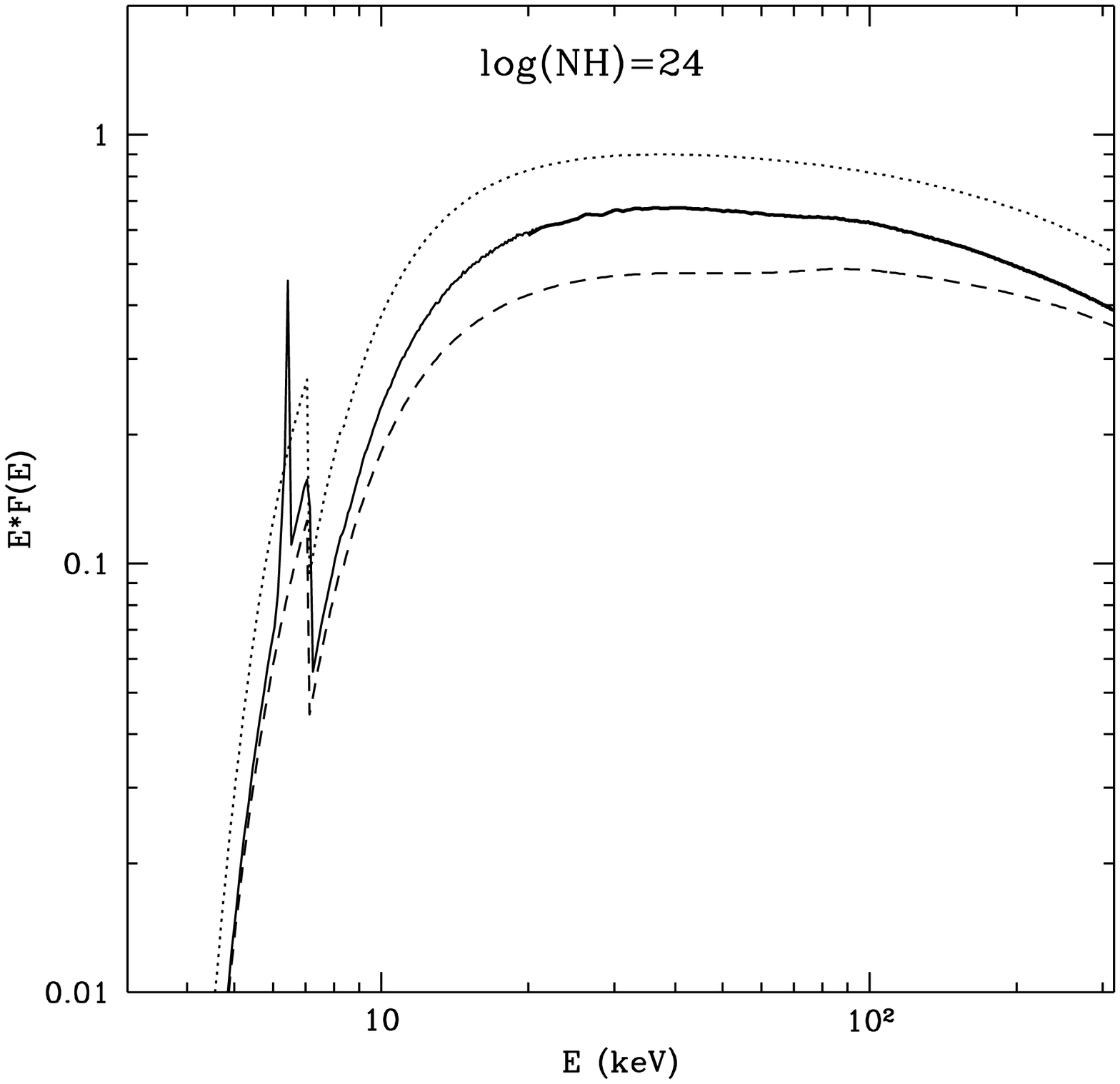, height=12.cm}
\caption{ As for Figure 1, but for N$_{\rm H}$=$10^{24}$
cm$^{-2}$.}
\end{figure}

\begin{figure}
\epsfig{file=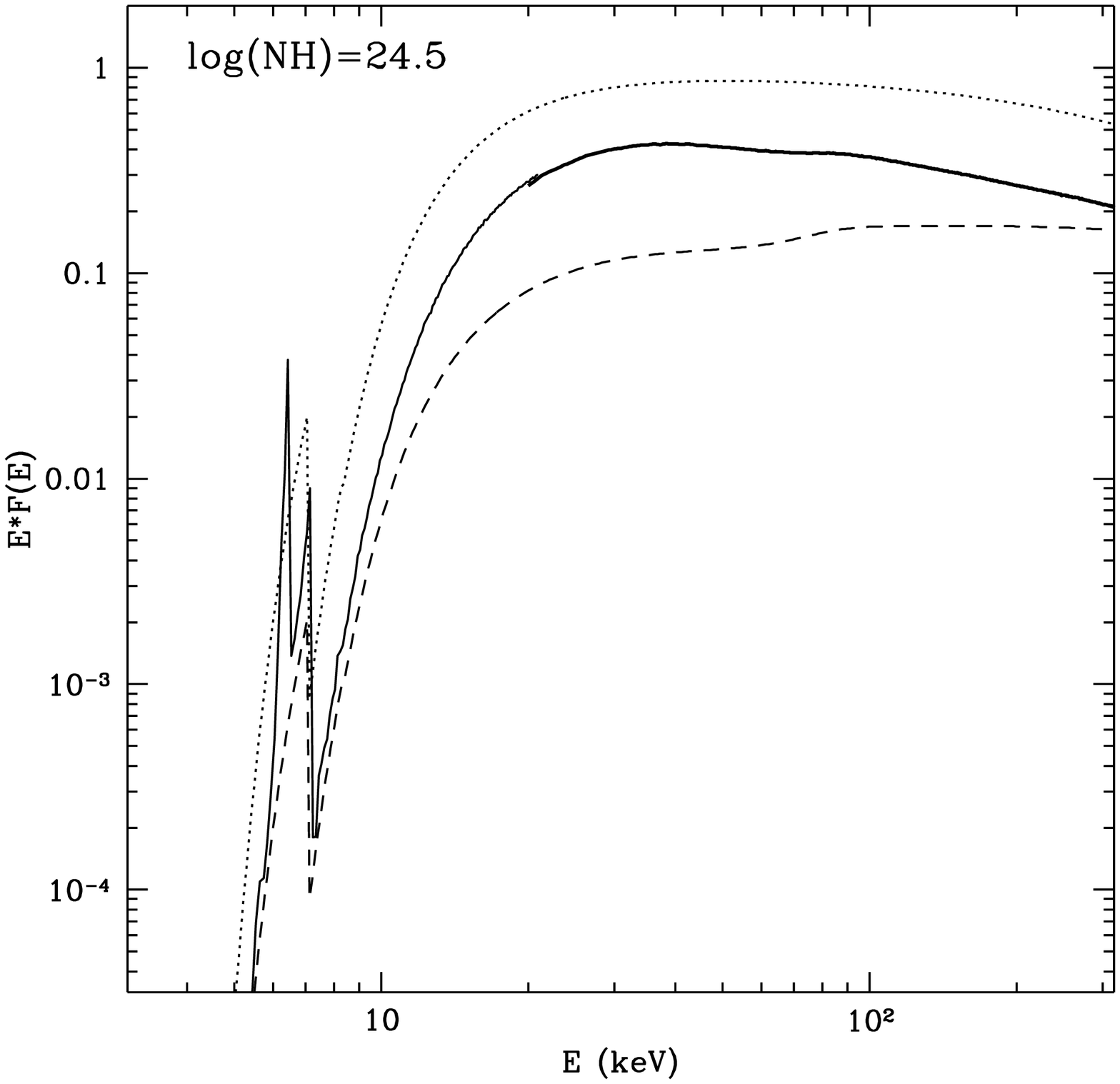, height=12.cm}
\caption{ As for Figure 1, but for N$_{\rm H}$=$3\times10^{24}$
cm$^{-2}$.}
\end{figure}

\begin{figure}
\epsfig{file=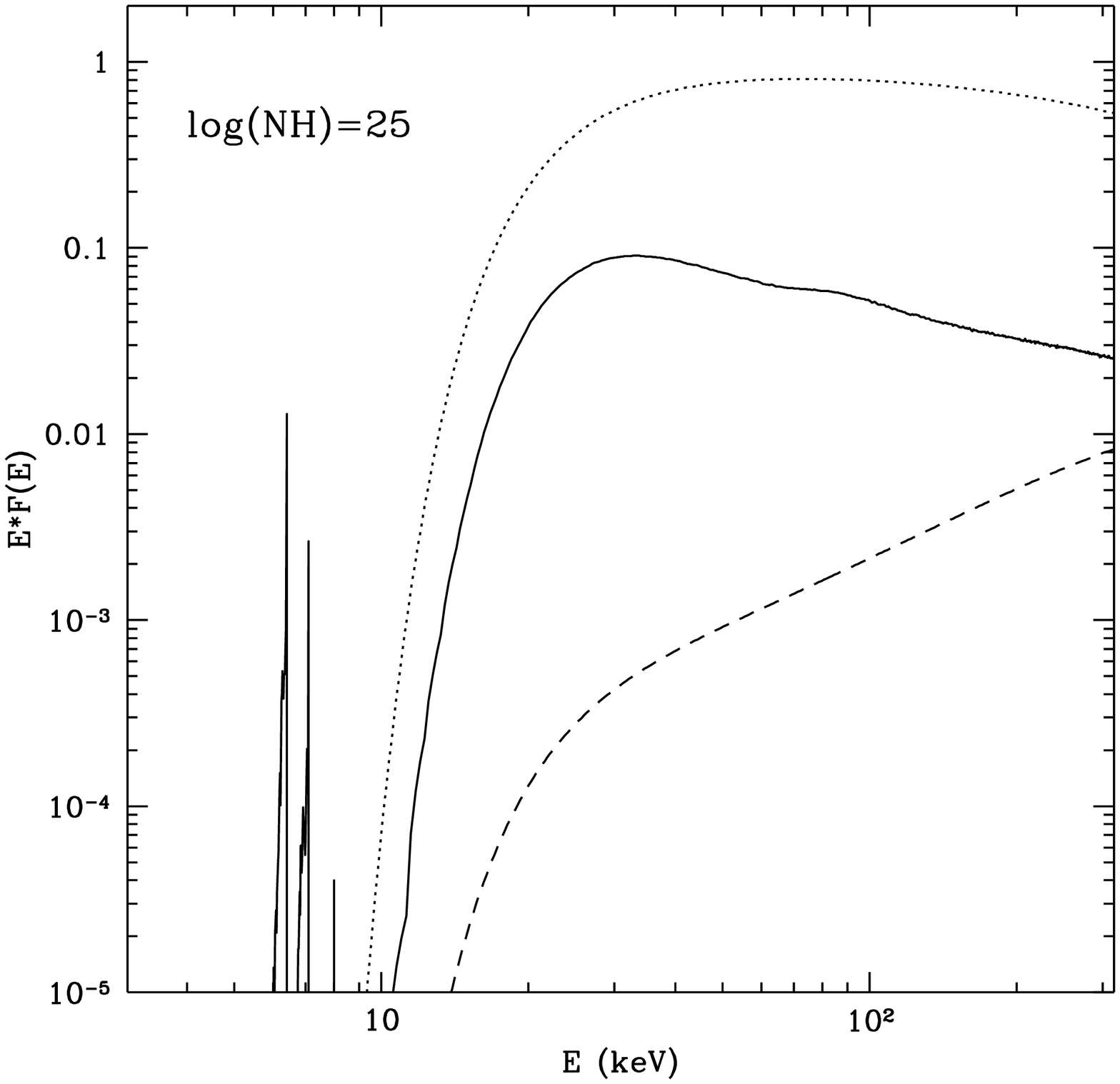, height=12.cm}
\caption{ As for Figure 1, but for N$_{\rm H}$=$10^{25}$
cm$^{-2}$.}
\end{figure}

\section{Applications. I. The Circinus Galaxy}

As a first application, let us discuss the case of the
Circinus Galaxy. Matt et al. 
(1999) analyzed the BeppoSAX observation of this source and found a 
clear excess in hard X--rays (i.e. in the PDS instrument)
with respect to the best fit medium energy (i.e. LECS and MECS) spectrum 
(which, in turn, was in good agreement with the ASCA result, Matt et al.
1996). The excess is best explained assuming that 
the nuclear emission is piercing through 
material with moderate Compton--thick (i.e. between $10^{24}$ and 
$10^{25}$ cm$^{-2}$, see previous section) column
density. Compton scattering must therefore be taken into
account in modeling the emerging spectrum, 
not only in absorption but also in emission, as
there is clear evidence of large amount of reflection too, suggesting 
a fairly large solid angle subtended by the cold matter to the primary
source. The fit with the model described here yields 
the parameters of the transmitted component
reported in Table~1 (model 1). Model 2 in the same table
refers to the fit with a pure absorption model (photelectric plus
Compton). Both models
are statistically acceptable (reduced $\chi^2\sim$1), but the differences 
in the best fit parameters are significant, leading to dramatically
different (i.e. two orders of magnitude) X--ray nuclear luminosities. 

\begin{table}
\vspace{0.15in}
\caption{Parameters of the best fit models. Model 1 includes Compton
scattering, model 2 only absorption. The primary spectrum is a power
law with photon index $\Gamma$ and a high energy cut--off; E$_{\rm C}$
is the corresponding e--folding energy. A is the density flux at 1 keV.}
\begin{tabular}{lcc}
\hline
\hline
~ & 1 & 2 \cr
\hline 
~ & ~ & ~\cr
N$_{\rm H,1}$~(10$^{24}$~cm$^{-2}$) & 4.3 & 6.9 \cr
A~(ph~cm$^{-2}$~s$^{-1}$~keV$^{-1}$~at~1~keV) & 0.11  & 15.4 \cr 
$\Gamma$ & 1.56 & 1.58 \cr
E$_{\rm C}$~(keV) & 56 & 38 \cr
L(2-10~keV)~(erg~s$^{-1}$) & 10$^{42}$ & 1.5$\times$10$^{44}$ \cr
~ & ~ & ~\cr
\hline
\hline
\end{tabular}
\vspace{0.15in}
\end{table}

\section{Applications. II. The hard X--ray Background}

The origin of the thermal--like, $\sim$40 keV 
spectrum (Marshall et al. 1980)
of the hard Cosmic X--ray background (XRB) has remained 
unexplained for many years. In 1989, Setti \& Woltjer 
proposed an explanation in terms of 
a mixture of obscured (i.e. Seyfert 2s) and unobscured 
(i.e. Seyfert 1s) AGN. Following this idea, 
many authors developed synthesis
models for the XRB (e.g. Madau, Ghisellini \& Fabian 1993, 1994;
Matt \& Fabian 1994; Comastri et al. 1995), and nowadays this explanation
is widely considered as basically correct. 

\begin{figure}
\epsfig{file=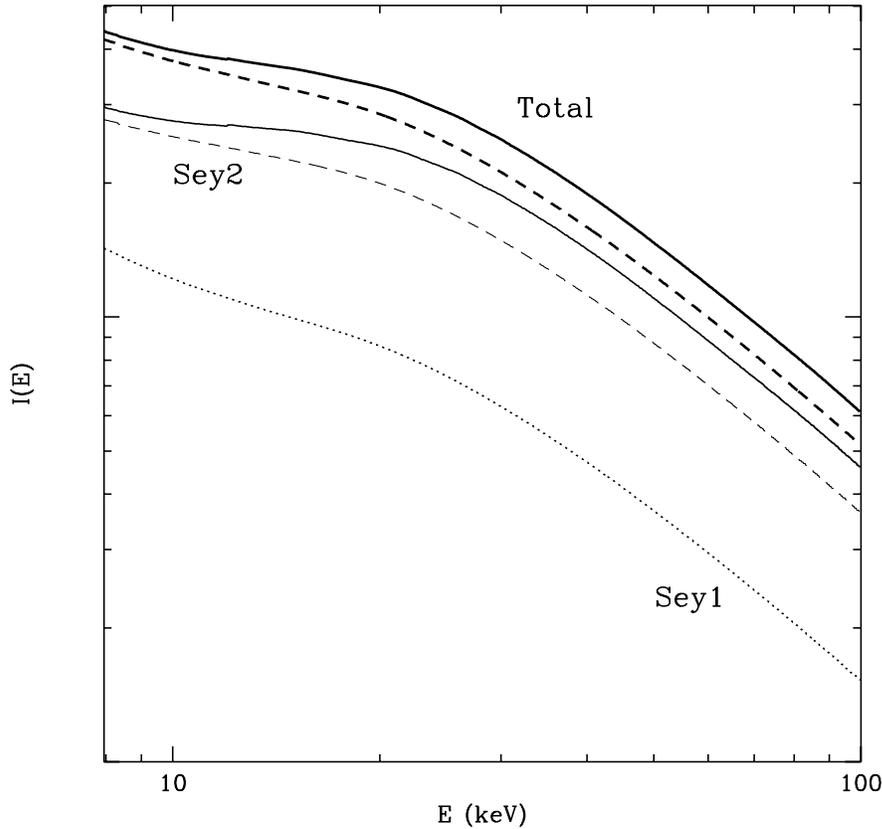, height=12.cm}
\caption{The integrated local spectrum of Seyfert 2s (with, solid line, 
and without, dashed line, the transmitted component), of Seyfert 1s (dotted
line) and total. See text for detail.}
\label{localsp}
\end{figure}

To model the spectrum of the XRB it is necessary
to include all the relevant ingredients, and a correct
transmission spectrum is one of them because, as remarked above,
Compton--thick sources are a significant fraction of all Seyfert 2s.
To our knowledge, out of the many papers devoted to fitting the XRB, 
the transmission component has been properly included 
only by Madau, Ghisellini \& Fabian (1994). 
Here we do it again, to highlight and discuss
the differences with models in which only absorption is included. 
In Fig.~\ref{localsp}, we show the integrated local spectrum of Seyfert 1
galaxies (dotted curve), of Seyfert 2
galaxies (lower dashed and solid curves) and of the sum of Seyfert 1 
and 2 galaxies (upper dashed and solid curves). 
The spectrum of Seyfert 1 galaxies is described by a power law with a
photon spectral index of 1.9 and an exponential cut--off with $e$-folding
energy of 400 keV; a Compton reflection component, corresponding to an
isotropically illuminated accretion disk observed at an inclination angle
of 60$^{\circ}$, is also included. According to unification models,
the spectrum of Seyfert 2 galaxies is assumed to be intrinsically 
identical to that of Seyfert 1s, but seen through obscuring matter.  
The solid lines in the figure refer 
to a synthesis model in which the transmitted 
component is included, while in the dashed ones only absorption 
is considered. Type 2 sources are assumed to outnumber type 1 sources
by a factor of 4, independently of the luminosity.
The adopted distribution of column densities of the absorbing
matter for the Seyfert 2s is: ${dN \over dLog(N_{\rm H})} \propto
Log(N_{\rm H})$, from 10$^{21}$ to 4$\times$10$^{25}$ cm$^{-2}$.
The fraction of Compton--thick sources
is then about 1/3, in agreement with the estimate of Maiolino et al. (1998).
The two total spectra differ significantly above 10 keV, the spectrum
including the transmission component being about 20\% higher at 30 keV.

\begin{figure}
\epsfig{file=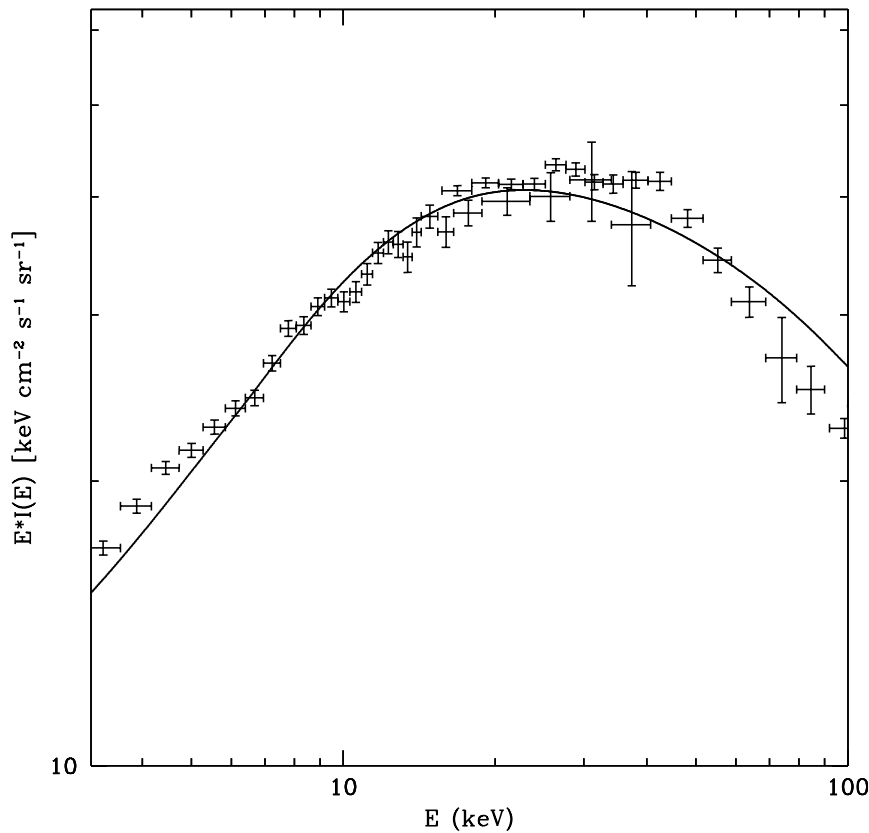, height=12.cm}
\caption{ Best fit spectrum to the XRB (data from HEAO-1, Marshall et al.
1980), using our model including the transmitted component.}
\label{fitsp}
\end{figure}

The best fit spectrum to the XRB (HEAO-1 data, Marshall et al. 1980),
obtained after evolving the local spectrum of Seyfert galaxies 
to cosmological distances, following Boyle et al. (1994),
is shown in Fig.\ref{fitsp}. 
Different descriptions of the pure luminosity evolution scenario
do not change significantly the results. The study
of both the spectral shape of the XRB and the source counts in 
different scenarios, 
including e.g. density evolution, is beyond the scope of the present work,
and is deferred to a forthcoming paper (Pompilio et al., in preparation).
Apart
from the highest energy part of the spectrum, where the fit is not very
good (suggesting either that an exponential cut--off is not a good
description of the spectrum of Seyfert galaxies, or that there is not
a universal value of such a parameter, as actually is emerging from BeppoSAX
observations: see e.g. Matt 1998),
and the lowest part (where contributions from other classes
of sources, like Clusters of Galaxies, may be relevant),
the agreement between the data and the model is acceptable.
The soft X--ray source counts are also well reproduced,
while the hard (5--10 keV) counts (Fiore et al., in preparation; Comastri 
et al. 1999) are somewhat underestimated, but still marginally consistent
with the data. The complete model and the
detailed analysis will be discussed elsewhere (Pompilio 1999; 
Pompilio et al., in preparation).

\begin{thebibliography}{}

\bibitem{} Boyle, B., Shanks, T., Georgantopoulos, I., Stewart, G.C.,
Griffiths, R.E. 1994, MNRAS, 271, 639

\bibitem{} Cappi, M., Bassani, L., Comastri, A., et al. 1999, A\&A, in press

\bibitem{} Comastri, A., Setti, G., Zamorani, G., Hasinger, G. 
1995, A\&A, 296, 1

\bibitem{} Comastri, A., Fiore, F., Giommi, P., La Franca, F.,
Elvis, M., Matt, G., Molendi, S., Perola G.C. 1999, Accepted for 
publication in Advances in Space Research, 
Proceedings of the 32nd Scientific Assembly of COSPAR (astro-ph/9902060)

\bibitem{} Done, C., Madejski, G.M., Smith, D.A. 1996, ApJ, 463, 63

\bibitem{} Iwasawa, K., Koyama, K., Awaki, H., Kunieda, H.,  Makishima, K.,
 Tsuru, T., Ohashi, T., Nakai, N. 1993, ApJ, 409, 155

\bibitem{} Madau, P., Ghisellini, G., Fabian, A.C. 1993, ApJ, 410, L7

\bibitem{} Madau, P., Ghisellini, G., Fabian, A.C. 1994, MNRAS, 270, 17

\bibitem{} Maiolino, R., Salvati, M., Bassani, L., Dadina, L., Della Ceca,
R., Matt, G., Risaliti, G., Zamorani, G. 1998, A\&A, 338, 781

\bibitem{} Marshall, F.E., Boldt, E.A., Holt, S.S., Miller R.B., 
Mushotzky R.F., Rose, L.A.,  Rothschild R.E., Serlemitsos P.J. 1980, ApJ,
235, 4

\bibitem{} Matt, G. 1998, to appear in ``High Energy Processes in accreting
black holes", J. Poutanen \& R. Svensson (eds) (astro-ph/9811053)

\bibitem{} Matt, G., Perola, G.C., Piro, L. 1991, A\&A, 247, 25

\bibitem{} Matt, G., Fabian, A.C. 1994, MNRAS, 267, 187

\bibitem{}  Matt, G.,  Fiore, F., Perola, G.C., Fink, H.H., Grandi, P.,
Matsuoka, M., Oliva, E., Salvati, M. 1996, MNRAS, 208, 253

\bibitem{} Matt, G., Guainazzi, M., Frontera, F., et al. 1997, A\&A,
325, L13

\bibitem{} Matt, G., Guainazzi, M., Maiolino, R., et al. 1999, A\&A,
341, L39

\bibitem{} Morrison, R., McCammon, D. 1983, ApJ, 270, 119

\bibitem{} Pompilio, F., 1999, Tesi di Laurea, Univ. Roma Tre


\bibitem{} Setti, G., Woltjer, L. 1989, A\&A, 224, L21

\bibitem{} Yaqoob, T. 1997, ApJ, 479, 184

\end{thebibliography}
\end{document}